# Superconductivity, Charge Orderings and Phase Separations in Systems with Local Electron Pairing

S. ROBASZKIEWICZ AND G. PAWŁOWSKI

*Institute of Physics, A. Mickiewicz University,*
*Umultowska 85, 61-614 Poznań, Poland*

## Abstract

We study two effective models developed for description of superconductors with short-coherence length: (i) the extended Hubbard model with on-site attraction and intersite repulsion, (ii) the model of hard-core charged bosons on a lattice. The analysis is concentrated on the problem of phase separations and competition between superconductivity (SS) and charge-density-wave (CDW) orderings. The phase diagrams of the systems are shown to consist of at least seven different states, including 3 types of phase separated (PS) states: CDW-SS (PS1), CDW-normal (PS2) and the state of electron droplets (PS3). By taking into account the PS states and the effects of longer-range density-density interactions (beyond nearest neighbours) our work substantially generalizes and modifies the conclusions of previous works concerning the models considered.

# 1 Introduction

High $T_C$ superconductors (HTS) (cuprates, barium bismuthates and alcali fullerenes) and several other nonconventional superconducting materials (e.g. the Chevrel phases) belong to a unique group of extreme type II superconductors with Ginzburg-Landau parameter $\kappa \gg 1$. They generally exhibit low carrier density, a small value of the Fermi energy ($E_F \leq 0.1 \div 0.3$ eV), a short coherence length $\xi_0$ and $\xi_0 k_F \approx 1 \div 10$ which indicates that in contrast to a weak coupling BCS theory the size of a pair can be of an order of the interparticle distance (or even a lattice constant) [1-5].

Detailed μSR studies show that all these systems have their $T_C$ proportional to $T_F$ (the Fermi temperature) or $T_B$ ( the Bose-Einstein condensation temperature) with $T_B \approx (3 \div 30) T_C$ and $T_F \approx (10 \div 100) T_C$ [3]. Moreover, for many classes of HTS strong empirical evidence has been found for the universal 3D X-Y critical behaviour as well as several universal trends have been established in the following dependences: Tc versus condensate density dependence, the $T_C$ dependence of the pressure and the isotope effect coefficients and the temperature dependence of the penetration depth [4, 5].

All these universal features and trends are certainly consistent with short-range, almost unretarded effective interaction responsible for pairing and they indeed support the models with local nonretarded attraction [1]. In our work we will study the effective models of this type:

(i) the extended Hubbard model (EHM) with local attraction (U < 0) and longer-range repulsion ($W_{ij}$):

$$\hat{H}_I = \sum_{ij\sigma} t_{ij} c^+_{i\sigma} c_{j\sigma} + U \sum_i n_{i\uparrow} n_{i\downarrow} + \frac{1}{2} \sum_{ij} W_{ij} n_i n_j - \mu \sum_i n_i \quad , \tag{1}$$

$$n = \frac{N_e}{N} = \frac{1}{N} \sum_i <n_i> \quad , \qquad n \in [0, 2] \quad ,$$

where $t_{ij}$ denotes the electron transfer integral from site i to j, $n_i = n_{i\uparrow} + n_{i\downarrow}$, $n_{i\sigma} = c^+_{i\sigma} c_{i\sigma}$, and μ is the chemical potential, $n = N_e/N$ is the number of particles per lattice site.

(ii) the model of hard-core charged bosons on a lattice:

$$\hat{H}_{II} = -\sum_{ij} J_{ij} b^+_i b_j + \sum_{ij} K_{ij} \hat{n}_i \hat{n}_j - \bar{\mu} \sum_i \hat{n}_i \quad , \tag{2}$$

{$b_{ij}^+$} are the Pauli operators, $\hat{n}_i = b_i^+ b_i$, $\bar{n} = \frac{N_b}{N} = \frac{1}{N} \sum_i <\hat{n}_i>$, $J_{ij}$ is the effective transfer integral of electron pairs (hard core bosons with charge 2e) from site i to j, $K_{ij}$ is the intersite density interaction between the pairs, $\bar{n}$ is the boson concentration per site ($2\bar{n} = n \in [0, 2]$).

The model (1) is the conceptually simplest lattice model to display a crossover from BCS-like to local pair (composite bosons) superconductivity with increasing local attraction as well as to describe the CDW orderings in narrow band materials. It has been considered an effective model for barium bismuthates [1, 6, 7], fullerides [8-10], the cuprates [1, 11-13], the Chevrel phases and for various classes of systems with alternating valence [1].

Eq(2) is the minimum Hamiltonian for the local pair models of superconductivity. This model is appropriate for description of the large $|U|/t$ limit of the model (1) [1, 14] and in such a case:

$$J_{ij} = 2t_{ij}^2/|U|, \qquad K_{ij} = J_{ij} + 2W_{ij}, \qquad \bar{n} = n/2, \qquad \bar{\mu} = 2(\mu + \frac{1}{2}|U| + \frac{1}{2}\sum_j W_{ij}). \qquad (3)$$

It also covers the case of the intersite pairing, large bipolarons and magnetic bipolarons [1, 2, 12].

For a review of earlier works concerning thermodynamics, electromagnetic properties and critical behaviour of the models (1) and (2), and for a detailed summary of theoretical and experimental arguments in favour of the local electron pairing description of the extreme type II superconductors, we refer the Reader to Refs [1, 2, 12].

Here, we will concentrate on the intriguing problem of phase separation and the competiton between superconductivity and charge orderings. We will present new results concerning 1) the evolution of the phase diagrams as a function of particle concentration and the interaction parameters and 2) the stability conditions of various phases possible in the system. By taking into account the phase separated states and the effects of longer-range density-density interaction our work substantially modifies and generalizes the conclusions of previous works concerning the phase diagrams of the models considered [1, 6, 7, 14-18].

In Table 1 we have summarized the types of ordered states in the systems considered, and defined the long-range order parameters describing each of these states.

## 2. The Extended Hubbard model.

Mutual stability of the homogeneous states (SS vs CDW vs M vs NO) in the model (1) has been considered in detail in Ref. [16-18], for the case of intersite interactions $W_{ij}$ restricted to nearest-neighbours (nn). In our analysis of the phase diagrams of this model, including the phase separated states PS1, PS2 and PS3 and the effects of $W_{ij}$ beyond nn, we will use, as in earlier works, the broken symmetry Hartree-Fock approach (HFA).

Free energies of the homogeneous pure phases (SS, CDW, NO) are derived in a standard way (see below and the Appendix A) and the phase boundary between the SS and the M phase is determined within RPA-HFA, from the softening of the collective mode $\omega_q$ in the SS phase at $\mathbf{q} = \mathbf{Q} = \left\{\dfrac{\pi}{a}, \dfrac{\pi}{a}, \dfrac{\pi}{a}\right\}$, i.e. from $\omega_Q = 0$ [17, 18].

The energies of the phase separated states are calculated from the expression

$$E_{ps} = m\, E(n_+) + (1 - m)\, E(n_-) ,\tag{4}$$

where $E(n_\pm)$ are the values of $E = <H>/N$ at $n = n_\pm$ corresponding to the lowest energy homogeneous solution, m is a fraction of the system with a charge density $n_+$, $1 - m$ is a fraction with a density $n_-$ ($n_+ > n_-$),

$$m n_+ + (1 - m) n_- = n \tag{5}$$

and

$$\frac{\partial E(n_-)}{\partial n_-} = \frac{E(n_+) - E(n_-)}{n_+ - n_-} \tag{6}$$

(Maxwell construction).

For PS1 and PS2 $n_+ = 1$, whereas for PS3 $n_+ = 2 - n_-$.

Taking into account (5) and (6), Eq (4) can be written as:

$$E_{ps} = E(n_+) - (n_+ - n) \frac{\partial E(n_-)}{\partial n_-} . \tag{7}$$

### 2.1. Basic formulation and selfconsistent equations

Within the framework of broken symmetry Hartree-Fock approach the mean-field Hamiltonian including the charge-density-wave (CDW) ordering and the singlet superconductivity (on-site pairing) is given by

$$H_0 = \sum_{k\sigma} A_k^\sigma c_{k\sigma}^+ c_{k\sigma} - \frac{1}{2} \sum_{k\sigma} (\Delta_c\, c_{k\sigma}^+ c_{k+Q\,\sigma} + \text{h.c.})$$
$$- \sum_k (\Delta_o\, c_{k\uparrow}^+ c_{-k\downarrow}^+ + \text{h.c.}) + \sum_k (\Delta_Q\, c_{k+Q\uparrow}^+ c_{-k\downarrow}^+ + \text{h.c.}) + C , \tag{8}$$

where

$$A_k^\sigma = \varepsilon_k - \frac{1}{N} \sum_q W_{k-q} <c_{q\sigma}^+ c_{q\sigma}> - \tilde{\mu} , \qquad \tilde{\mu} = \mu - n\left(-\frac{|U|}{2} + W_o\right) ,$$

$\varepsilon_k$ and $W_k$ are the Fourier transforms of $t_{ij}$ and $W_{ij}$, respectively.

C is a constant arising from the Gorkov-type factorization:

$$C = N\left[(|U| - 2W_o)\frac{n^2}{4} - (|U| - 2W_Q)\frac{n_Q^2}{4} - \frac{|\Delta_o|^2}{|U|} - \frac{|\Delta_Q|^2}{|U|}\right] + \frac{1}{2N}\sum_{kq\sigma} W_{k-q} <c^+_{k\sigma} c_{k\sigma}><c^+_{q\sigma} c_{q\sigma}> .$$

(9)

The CDW gap parameter is defined by

$$\Delta_c = (|U| - 2W_Q)\frac{n_Q}{2} ,$$

where $n_Q = \frac{1}{N}\sum_{k\sigma} <c^+_{k+Q\sigma} c_{k\sigma}> ,$

and **Q** is half the smallest reciprocal lattice vector ($\mathbf{Q} = \frac{\pi}{a}(1, 1, 1)$ for the SC lattice).

The superconducting order parameters are

$$\Delta_o = \frac{1}{N}|U|\sum_k <c_{-k\downarrow} c_{k\uparrow}> = |U| x_o , \qquad \Delta_Q = \frac{1}{N}|U|\sum_k <c_{k\downarrow} c_{-k+Q\uparrow}> = |U| x_Q ,$$

$\Delta_o$ and $\Delta_Q$ determine the uniform and the alternating component of superconducting ordering, respectively, and $\Delta_Q$ is called the $\eta$-paring order parameter [19]. With $\Delta_Q$ taken into account in $H_o$ our approach is more general than that used in previous works [1, 7, 10, 16, 20], where it was assumed that $\Delta_Q = 0$.

$H_o$ can be diagonalized with the use of either the Green's-function method or the equation-of-motion approach. In the case of alternating lattices (i.e. for $\varepsilon_k = -\varepsilon_{k+Q}$) we obtain the following eigenvalues of $H_o$:

$$\omega^k_{1,4} = \pm A_+(\mathbf{k}), \quad \omega^k_{2,3} = \pm A_-(\mathbf{k}),$$

(10)

where

$$A_\pm(\mathbf{k}) = \left((\bar{\varepsilon}_\mathbf{k})^2 + \tilde{\mu}^2 + \Delta_c^2 + \Delta_o^2 + \Delta_Q^2 \pm 2\sqrt{(\bar{\varepsilon}_\mathbf{k})^2(\tilde{\mu} + \Delta_Q^2) + \tilde{\mu}^2 \Delta_c^2 + \Delta_c^2 \Delta_Q^2 + 2\tilde{\mu}\Delta_c\Delta_o\Delta_Q}\right)^{1/2} ,$$

(11)

$$\bar{\varepsilon}_\mathbf{k} = \varepsilon_\mathbf{k} - p\frac{W_{1k}}{z} = \varepsilon_\mathbf{k}(1 - \frac{pW_1}{z\,t}) ,$$

(12)

$$\varepsilon_\mathbf{k} = t\gamma_\mathbf{k}, \quad \gamma_\mathbf{k} = \frac{1}{z}\sum_\delta e^{ik\delta} , \quad (\gamma_\kappa = 2(\cos k_x + \cos k_y + \cos k_z) \text{ for SC lattice}),$$

$$p = \frac{1}{N}\sum_k \gamma_\mathbf{k} <c^+_{k\sigma} c_{k\sigma}> .$$

In Eqs (10-12) we have taken into account only the Fock term p connected with the nearest-neighbour $W_{ij}$ ($W_1$). $\delta$ being the nn lattice vector and z is the number of nn.

In terms of the eigenstates of $H_o$, the expression for the free energy

$$F_o = -\frac{1}{\beta}\ln \text{Tr} \exp(-\beta H_o) + \mu N_e$$

is obtained as

$$\frac{F_o}{N} = \tilde{\mu}(n-1) + \frac{1}{4}\left[-|U| + 2W_o\right]n^2 + \frac{\Delta_c^2}{|U| - 2W_Q} + \frac{W_1}{z}p^2 +$$
$$+ \frac{|\Delta_o|^2}{|U|} + \frac{|\Delta_Q|^2}{|U|} - \frac{1}{\beta N}\sum_k \ln\left[4\cosh\frac{\beta A_k^+}{2}\cosh\frac{\beta A_k^-}{2}\right], \tag{13}$$

where the optimum values of $\tilde{\mu}$, p and the order parameters are given by the set of equations

$$\frac{\partial F_o}{\partial \tilde{\mu}} = 0, \quad \frac{\partial F_o}{\partial p} = 0, \quad \frac{\partial F_o}{\partial \Delta_c} = 0, \quad \frac{\partial F_o}{\partial \Delta_o} = 0, \quad \frac{\partial F_o}{\partial \Delta_Q} = 0. \tag{14}$$

For $W_{ij}$ restricted to nn and next nn:

$$W_o = zW_1 + z_2W_2, \quad W_Q = -zW_1 + z_2W_2, \tag{15}$$

whereas for $W_{ij}$ having the long-range part:

$$W_o = zW_1 + W_{LR}, \quad W_Q = -zW_1 + W_{LR}^Q, \tag{16}$$

where $W_{LR} = \sum_{n=2}^{\infty} z_n W_n$, $W_{LR}^Q = \sum_{n=2}^{\infty} z_n W_n (-1)^n$.

In the case when $\Delta_Q = 0$, $p = 0$ and $W_{ij}$ are restricted to nearest neighbours, Eqs (10-14) reduce to those derived in Ref. [16].

Eqs (13), (14) describe the mixed CDW-SS phase of the model ($\Delta_c \neq 0$, $\Delta_o \neq 0$, $\Delta_Q \neq 0$, cf. table 1). The corresponding equations for the homogeneous SS, CDW and NO phases have the following forms:

The SS phase

$$\frac{F_o}{N} = \tilde{\mu}(n-1) + \frac{1}{4}\left[-|U| + 2W_o\right]n^2 + \frac{W_1}{z}p^2 + \frac{|\Delta_o|^2}{|U|} - \frac{2}{\beta N}\sum_k \ln\left[2\cosh\frac{\beta E_k}{2}\right], \tag{17}$$

where

$$E_k = \left(\tilde{\varepsilon}_k^2 + \Delta_o^2\right)^{1/2},$$
$$\tilde{\varepsilon}_k = \varepsilon_k - p\frac{W_1 k}{z} - \tilde{\mu}.$$

Minimizing $F_o$ with respect to $\tilde{\mu}$, p, and $\Delta_o$ one obtains:

$$\Delta_o = |U|\frac{1}{N}\sum_k \Delta_o F_k, \quad p = -\frac{1}{N}\sum_k \tilde{\varepsilon}_k \gamma_k F_k, \quad n - 1 = \frac{-2}{N}\sum_k \tilde{\varepsilon}_k F_k, \tag{18}$$

where

$$F_k = (2E_k)^{-1}\tanh\frac{\beta E_k}{2}.$$

## The CDW phase

$$\frac{F_o}{N} = \tilde{\mu}(n-1) + \frac{1}{4}[-|U| + 2W_o]n^2 + \frac{W_1}{z}p^2 + \Delta_c^2/(|U| - 2W_Q) - \frac{1}{\beta N}\sum_{k\alpha}\ln[2\cosh\frac{\beta E_k^\alpha}{2}],$$

(19)

where

$$E_k^\alpha = -\tilde{\mu} + \alpha\sqrt{\overline{\varepsilon_k^2} + \Delta_c^2}, \quad \alpha = \pm$$

$$\Delta_c = \frac{|U| - 2W_Q}{2N}\Delta_c\sum_{k\alpha}\frac{\alpha\tanh\beta\frac{E_k^\alpha}{2}}{2\sqrt{\overline{\varepsilon_k^2} + \Delta_c^2}}, \quad n - 1 = -\frac{1}{2N}\sum_{k\alpha}\tanh(\beta\frac{E_k^\alpha}{2}),$$

$$p = -\frac{1}{2N}\sum_{k\alpha}\gamma_k\,\overline{\varepsilon}_k\,\frac{\alpha\tanh\beta\frac{E_k^\alpha}{2}}{2\sqrt{\overline{\varepsilon_k^2} + \Delta_c^2}}.$$

## The NO phase

$$\frac{F_o}{N} = \tilde{\mu}(n-1) + \frac{1}{4}[-|U| + 2W_o]n^2 + \frac{W_1}{z}p^2 - \frac{2}{\beta N}\sum_k\ln[2\cosh\frac{\beta\tilde{\varepsilon}_k}{2}],$$

(20)

where

$$n - 1 = -\frac{2}{N}\sum_k\tilde{\varepsilon}_k\,F_k(T),$$

$$p = -\frac{1}{N}\sum_k\gamma_k\,\tilde{\varepsilon}_k\,F_k(T), \quad F_k(T) = (2\tilde{\varepsilon}_k)^{-1}\tanh(\beta\frac{\tilde{\varepsilon}_k}{2}).$$

The condition determining the phase boundary between the SS and the M phase can be derived in the simplest way within the RPA-HFA theory, from the softening of the collective excitation mode $\omega_q$ in the SS phase at $\mathbf{q} = \left\{\frac{\pi}{a},\frac{\pi}{a},\frac{\pi}{a}\right\} = \mathbf{Q}$, i.e. from $\omega_Q = 0$. For $T = 0$ it has been done in Ref. [17] and we only quote the resulting equations:

$$|U| - 2W_Q = \frac{1}{|U|} - \frac{\frac{1}{2}\tilde{\mu}\,S_Q}{1 + \frac{|U|\Delta_o^2\,S_Q}{2\tilde{\mu} + |U|(1-n)}},$$

(21)

where $\Delta_o$ and $\tilde{\mu}$ are given by a solution of the set of the self-consistent equations

$$\Delta_o = \frac{|U|\Delta_o}{2N}\sum_k E_k^{-1},$$

$$n - 1 = \frac{1}{N}\sum_k\tilde{\varepsilon}_k\,E_k^{-1}, \quad p = -\frac{1}{2N}\sum_k\tilde{\varepsilon}_k\,\gamma_k\,E_k^{-1},$$

(22)

and

$$E_K = (\tilde{\varepsilon}_k^2 + \Delta_o^2), \quad S_Q = P \sum_k \frac{1}{\tilde{\varepsilon}_k E_k} \quad ,$$

P denotes the principal value.

In our work we concentrate on the studies of the ground state properties. Discussion of the results for finite temperatures will be given in a separate paper.

We have performed a detailed analytical and numerical analysis of the derived equations at T = 0 ($\beta \to \infty$), and examined the phase diagrams of the system as a function of the interaction parameters and the electron concentration.
In the calculations we have made use of the rectangular density of states (DOS)

$$N(\varepsilon) = 1/2 D \quad , \quad \text{for } D < |\varepsilon_k|, \tag{23}$$
$$= 0 \quad , \quad \text{otherwise.}$$

with D denoting the effective half-bandwidth.
Several analytical results concerning the ground state characteristics of the system, which can be derived for the SS, CDW and NO phases as well as for the PS states are collected in the Appendix.

To find the stable state and construct the phase diagrams we have compared the energies $E_o = F_o(\beta \to \infty)/N$ of SS, CDW, NO, PS1, PS2, PS3 states, choosing the solution which yield the lowest value of $E_o$ and we calculated the phase boundary between the M phase and the SS phase from Eqs (21), (22).

## 2.2 Phase diagrams

Depending on the particle concentration n (0 < n < 2) and the interaction parameters the phase diagrams of the system considered are found to consist of at least seven different states, i.e., normal (NO) (which appears only at T > 0), singlet superconducting (SS), charge ordered (CDW), homogeneous CDW-SS (M), CDW-SS phase separated (PS1), CDW-NO phase separated (PS2) and the state of particle droplets (PS3) (cf. Table 1).

The evolution of the ground state phase diagrams (for model (1)), as a function of particle concentration and interactions is presented in Figs 1-4.

A. <u>The case of nn interactions</u>

In Fig.1 we show the ground state phase diagrams of model (1) for $W_{ij}$ restricted to nn, plotted for various values of $|U|/D$. For repulsive nn $W_{ij}$ the ground state consists of the CDW phase for n = 1, the PS1 state for $1 > n > n_c$ and $1 < n < \bar{n}_c$ and the SS phase for $n < n_c$ and for $n > \bar{n}_c$, where $n_c$ and $\bar{n}_c = 2 - n_c$ are the critical concentrations dependent on $|U|/D$ and $W/D$, at which the first order transitions between the PS1 state and the SS phase take place. With increasing $|U|/D$ and $W/D$ the region of electron concentration in which the PS1 state is stable grows at the expense of the region of the stable SS phase. For weak attractive $W_{ij}$, the ground state is SS for all electron concentrations, and increasing $|W_{ij}|$ (W < 0) stabilizes the state of particle droplets (PS3).

In Fig.1 the dashed lines denote the range of existence of the M phase solutions for a given $|U|/D$. As we see, in the case considered the instability of the SS phase with respect to the M phase is preceded by a first order transition to the PS1 state and therefore it has no effect on the phase diagrams. However, the distance between the PS1 - SS boundary and the M - SS boundary is diminished with increasing $|U|/D$ and it tends to zero for $|U|/D \to \infty$. Such a result is in agreement with the strong attraction theory (see next section) which predicts that for large $|U|$ the ground state energies of the PS1 and M states are strictly degenerated if $W_{ij}$ is restricted to nn.

### B. The effects of longer-range interaction (beyond nn)

Fig. 2 shows the effects of the second neighbour density interactions $W_2$ on the phase diagram. A general tendency is that the repulsive (attractive) $W_2$ reduces (expands) the stability regions of both phase separated states PS1 and PS3.

The evolution of the boundary between the PS1 and the SS states as well as the boundary between the M and the SS phases with increasing repulsive $W_2$ is given in Figs 3a,b for several fixed values of $|n - 1|$. Both boundaries are shifted towards higher values of $W_o$ but they do not intersect with increasing $W_2$. Thus one can conclude that in the weak and intermediate coupling regime, i.e. for $|U|/2D \leq 1$, the homogeneous M phase remains unstable with respect to the PS1 state also in the presence of $W_2 > 0$.

The effects of the long-range repulsion $W_{LR}$ on the phase diagram are much more drastic. $W_{LR}$ tends to destabilize the PS1 state with respect to both the SS phase and the M phase. As we see from Figs. 4, even a relatively weak $W_{LR}$ can strongly reduce the range of stability of the PS1 state and allow the homogeneous M phase to develop in the high density limit. The PS1 remains stable only in a restricted region of small $W_o/D$ and $|n - 1|$, which is reduced with increasing $|U|/D$.

## 3. Hard core bosons on a lattice

In analysis of model (2) we have adopted the MFA variational approach [14, 21]. After diagonalization of the trial Hamiltonian, the free energy is obtained as:

$$F_o = -\frac{1}{\beta} \sum_m \ln[2\cosh\beta\Delta_m] + \frac{1}{2}\sum_{mm'}\left(J_{mm'}\psi_m^+ \psi_{m'}^- + h.c.\right) - \sum_{mm'}K_{mm'}\psi_m^z \psi_{m'}^z + \bar{\mu}(\bar{n} - \frac{1}{2})N$$
$$+ \text{const.} \tag{24}$$

where

$$\Delta_m = \left((\Omega_m^z)^2 + (\Omega_m^+)^2\right)^{1/2},$$
$$\Omega_m^z = -\sum_{m'}K_{mm'}\psi_{m'}^z + \frac{\bar{\mu}}{2},$$
$$\Omega_m^\pm = -\sum_{m'}J_{mm'}\psi_{m'}^\pm, \tag{25}$$
$$\psi_m^+ = <b_m^+> = (\psi_m^-)^+,$$
$$\psi_m^z = <n_m> - \frac{1}{2}. \tag{26}$$

From $\partial F_o / \partial \psi_m^{\pm,z} = 0$ one gets a set of equations determining $\psi_m^\alpha$ and $\bar{\mu}$

$$\psi_m^- = \frac{\Omega_m^-}{2\Delta_m}\text{th}(\beta\Delta_m), \quad \psi_m^z = \frac{\Omega_m^z}{2\Delta_m}\text{th}(\beta\Delta_m), \quad \frac{n-1}{2} = \frac{1}{N}\sum_m \psi_m^z, \quad (n = 2\bar{n}). \tag{27}$$

Assuming the existence of two interpenetrating lattices A, B, and restricting the analysis to the two-sublattice solutions we define the order parameters

$$x_o = \frac{1}{N}\sum_m \langle b_m^+ \rangle = \frac{\psi_A^+ + \psi_B^+}{2},$$

$$x_Q = \frac{1}{N}\sum_m \langle b_m^+ \rangle \exp(iQ\mathbf{R_m}) = \frac{\psi_B^+ - \psi_A^+}{2},$$

$$n_Q = \frac{1}{N}\sum_m \langle n_m \rangle \exp(iQ\mathbf{R_m}) = \frac{\psi_B^z - \psi_A^z}{2},$$

which describe the SS phase ($x_o \neq 0$), the CDW phase ($n_Q \neq 0$) and the M phase ($x_o \neq 0$, $x_Q \neq 0$, $n_Q \neq 0$).

From Eqs (24), (27) the ground state energies $E_o = F_o$ ($\beta \to \infty$) for these particular phases and the NO phase are found to be ($n \leq 1$, $J_q$ and $K_q$ are the Fourier transforms of $J_{ij}$ and $K_{ij}$, respectively):

$$E_o^{SS} = 1/4 \ (K_o + J_o)(1 - n)^2 - J_o/4, \tag{28}$$

$$E_o^{NO} = K_o/4 \ (1 - n)^2 \tag{29}$$

$$E_o^{M} = E_o^{SS} - 1/4 \ (\sqrt{-K_Q + J_Q} \ (1 - n) - \sqrt{-K_Q + J_o}) \tag{30}$$

The M phase solutions exist if $K_Q + J_Q < 0$ and $K_Q + J_o < 0$ provided that $n_c < n < 1$ (or $1 < n < 2-n_c$), where

$$n_c = 1 - \left(\frac{K_Q + J_o}{K_Q + J_Q}\right)^{1/2}, \tag{31}$$

and in such a case $E_o^M < E_o^{SS}$.

The CDW phase can appear only for $n = 1$, provided that $K_Q + J_Q < 0$ and

$K_Q + J_o < 0$, and

$$E_o^{CDW}(n=1) = 1/4 \ K_Q, \tag{32}$$

Using the Eqs (4)-(7) and (28), (29), (32) the energies of the phase separated states PS1, PS2 and PS3 (cf. Table 1) are derived as

$$E_o^{PS1} = 1/4 \ \{K_Q + 2(1 - n)\ [-(K_Q + J_o)(K_o + J_o)]^{1/2}\}, \tag{33}$$

$$E_o^{PS2} = 1/4 \ \{K_Q + (1 - n)(-K_Q + K_o)\}, \tag{34}$$

$$E_o^{PS3} = 1/4 \ K_O. \tag{35}$$

The PS1 solutions exist if $K_Q + J_Q < 0$ and $K_Q + J_o > 0$ provided that $n_p < n < 1$ (or $1 < n < 2-n_p$), where

$$n_p = 1 - \left(\frac{-(K_Q + J_o)}{K_o + J_o}\right)^{1/2}$$

and in this case $E_o^{PS1} < E_o^{SS}$.

Comparing Eqs (30) and (33) one can easily find that in the case of nn interactions, (i.e. for $K_Q = -K_o$ and $J_Q = -J_o$), $E_o^M = E_o^{PS1}$. Any longer range interaction $K_{ij}$ removes this degeneracy favouring the M or the PS1 state.

In Fig. 5 we show the phase diagrams of the model (2) including the nn and nnn density-density interactions (In this Figure we have used the denotations of Eq.(3): $K_{ij} = J_{ij} + 2W_{ij}$ for comparison with results of sec. 2).

For $K_{ij}$ restricted to nn (Fig 5a) the PS1 state is strictly degenerated with the M phase in the whole range of stability of both these states.
Repulsive $W_2$ destabilizes the PS1 state with respect to the M phase and reduces the range of stability of the PS3 state (Fig 5b). Attractive $W_2$ extends the stability regions of phase separated states at the cost of the homogeneous SS phase and eliminates the M phase (Fig 5c).

Notice that the PS1 state (superconducting) can be stable only for not too large nearest neighbour repulsion $W_1$. Above a critical value of $W_1$ (dependent on $W_2/J_o$) it is replaced by the PS2 state (nonsuperconducting), at an arbitrary particle concentration. The explicit expression for the PS1-PS2 borderline is $\dfrac{-(K_Q + K_o)}{2J_o} = 1$ .

## 4. Final remarks

Let us summarize our results:

1. In the systems considered at least 3 types of the phase separated states can develop besides the homogeneous phases. Two of them (PS1 and PS2) can appear in the case of **repulsive** n.n interaction $W_1 > 0$, the third one (PS3) - in the case of **attractive** interaction $W_1 < 0$ (compare Figs.1–3)

2. In the absence of longer-range interactions (beyond nn) the superconducting PS1 state (CDW-SS) is always more stable than the homogeneous M phase except $|U|/D \gg 1$, where they become energetically degenerated (compare Figs 1 and 3a).

3. The region of stability of PS1 state is reduced by both the next nn repulsion ($W_2 > 0$) and the long-range Coulomb interactions $W_{LR}$, but except for $|U|/D \gg 1$ it can be completely suppressed only by strong (unscreened) WLR (comp. Figs. 2, 3).

4. The most favorite conditions for the appearance of the M phase are found in the strong coupling limit ($|U|/D \gg 1$). In such a case an arbitrary weak $W_2$ (or $W_{LR}$) stabilizes this phase with respect to the PS1 (Fig. 3). On the contrary, in the weak coupling regime the M phase can develop only in the presence of substantial $W_{LR}$ (Fig. 2).

As we see the inclusion of the PS states and the effects of longer range Coulomb interactions into consideration substantially modifies the phase diagrams of the models considered obtained assuming only homogeneous phases.

In the phase separated states the system breaks into coexisting domains of two different charge densities $n_+$ and $n_-$. In real systems the sizes of the domains will be finite and determined by the long-range Coulomb repulsion and disorder effect (structural imperfections, disorder of doped ions, etc).

The properties of the system in the PS states can evolve with electron concentration. Let us take the PS1 state as an example: If the superconducting fraction of the system is large, (as it occurs near the border with th homogeneous SS phase, i.e. for $n_c \leq n$) the measured penetration depth will remain constant with increasing n (since $n_-$ does not depend on n). On the other hand, near the half-filling ($n \approx 1$) the SS fraction is strongly reduced and there will be only diluted SS domains ($n_- < 1$) in a semiconducting CDW background ($n_+ = 1$). When the SS domains do not percolate one should observe a partial Meissner effect but without zero resistance.

The present study can be extended in several directions. In the case of extended Hubbard model and large intersite repulsion, it would be important to generalize our HFA treatment by taking into account the additional variational parameters connected with the intersite (extended s-wave pairing) components of superconducting order (the averages: $<c^+_{i\uparrow} c^+_{j\downarrow}>$, $i \neq j$).

Our preliminary analysis and the results of ref. [22] concerning the homogeneous SS phase indicate that such a generalized approach is able to describe a complete suppression of on-site pairing by strong nearest neighbour repulsion in the weak and intermediate coupling limit ($|U|/D < 1$).

In the case of the hard-core bosons model a better treatment of the ground state and finite temperature thermodynamics including the collective excitation effects could be the self-consistent RPA-type approach. Till now only the properties of the homogeneous SS phase have been analysed within such an approach [1, 15, 23].

Further steps in the analysis of both models should also include the incommensurate CDW phases [18] as well as various types of striped phases [24] and "island type" phases [25], which can eventually appear in the presence of longer range interactions. These investigations and those concerning the effects of electron-lattice couplings (static strains) and structural disorder are left for future studies.

Among the materials for which the local electron paring has been either established or suggested the best candidates to exhibit the phase separation phenomena are the doped barium bismuthates ($BaPB_{1-x}Bi_xO_3$ and $Ba_{1-x}K_xBiO_3$) [1, 2, 6, 7, 26]. For these systems, being oxide perovskites, a very large dielectric constant strongly weakens the long-range Coulomb repulsion, which is the main factor preventing the phase separation [27].

There are several experimental observation suggesting a possibility of phase separation in these materials. In particular:

1) the gap in the optical spectra in insulating phase persisting as a pseudogap in the superconducting phase [1, 2] - indication for PS1 (SS-CDW) with local CDW ordering in domains,

2) anomalous magnetoresistance in the superconducting state [26, 28] and anomalous temperature dependences of critical current and critical magnetic field [28],

3) strong metastability and hysteresis effects observed near the boundary with semiconducting phase [6, 7, 26],

4) maximum $T_c$ located near the boundary with the CDW (or local CDW) phase [1, 2].

Further experimental studies which would help to test the theoretical predictions concerning the phase separation phenomena in this class of systems should include the search for local CDW ordering in domains (in both: the superconducting and semiconducting states) as well as detailed quantitative investigations of the Meissner effect and the penetration depth as a function of doping.

## ACKNOWLEDGEMENTS

We would like to thank R. Micnas and T. Kostyrko for many useful discussions. S. R. acknowledges the hospitality of the Institute for Scientific Interchange, Turin, Italy, where part of this work was discussed.

This work was supported by K.B.N. Poland, projects 2 P302 038 07 and 2PO3B05709.

## Appendix

Below we present explicit expressions concerning for the ground state energies and other T = 0 characteristics of the model (1) which can be derived using the rectangular density of states (Eq. (23)). The derivations for the homogeneous SS, CDW and NO phases are analogous to those performed in Ref. [16], therefore we quote only the final results:

<u>The SS phase:</u>

$$E_o^{ss} = -\frac{D}{2} n (2-n) \coth\frac{2D}{|U|} + \frac{1}{2} \widetilde{W}_o n^2, \tag{A1}$$

$$\widetilde{\mu} = (n-1) \coth\frac{2D}{|U|} ,$$

$$\Delta_o = |U| x_o = \sqrt{n(2-n)} \frac{D}{\sinh\frac{2D}{|U|}} ,$$

where

$$2\widetilde{W}_o = -|U| + 2 W_o .$$

<u>The CDW phase:</u>

$$E_o^{CDW}(n) = D\left(\frac{|n-1|}{\sinh\frac{2D}{2\widetilde{W}_Q}} - \frac{1}{2}[1-(n-1)^2]\coth\frac{2D}{2\widetilde{W}_Q}\right) + \frac{1}{2}\widetilde{W}_o n^2, \tag{A2}$$

$$\Delta_c^2 = \alpha^2 - (n-1)^2 D^2 ,$$

$$\widetilde{\mu} = \alpha, \quad \text{for } n \neq 1, \quad \widetilde{\mu} = 0, \quad \text{for } n = 1,$$

where

$$\alpha = D\left\{1 - |n-1|\cosh\frac{2D}{2\widetilde{W}_Q}\right\}\bigg/\sinh\frac{2D}{2\widetilde{W}_Q} ,$$

$$2\widetilde{W}_Q = -|U| + 2 W_Q .$$

<u>The NO phase:</u>

$$E_o^{NO} = -\frac{D}{2} n (2-n) + \frac{1}{2} \widetilde{W}_o n^2 , \tag{A3}$$

$$\widetilde{\mu} = (n-1) D .$$

In the above expressions we have neglected the contribution from the normal Fock term p, since its only effect appears as a renormalization of the bandwidth (D → D + $pW_1$) and it has little influence on the phase stability at T = 0.

The ground state energies of the phase separated states PS1, PS2 and PS3 are calculated using Eqs. (4) - (7) and taking for $E(n_\pm)$ the values of $E_o$ corresponding to the lowest energy homogeneous phase solution at $n = n_\pm$, i.e. $E_o^{SS}$(Eq(A1)),

$E_o^{CDW}$(Eq(A2)) or $E_o^{NO}$(Eq(A3)). The results obtained in such a way are the following:

<u>The PS1 state</u>

$n_+ = 1, \quad n_- = n_1 \leq 1,$

$E(n_+) = E(n_+=1) = E_{CDW}(n=1),$

$E(n_-) = E_{SS}(n_-).$

$$E_{PS1}(n) = -\frac{D}{2}\coth\frac{2D}{2\mathbb{W}_Q} + \frac{1}{2}\mathbb{W}_o + (1-n)\left(D(1-n_1)\coth\frac{2D}{|U|} - \mathbb{W}_o n_1\right), \tag{A4}$$

where

$$n_1 = 1 - \left(\frac{\coth\frac{2D}{2\mathbb{W}_Q} - \coth\frac{2D}{|U|}}{\coth\frac{2D}{|U|} + \frac{\mathbb{W}_o}{D}}\right)^{1/2}.$$

<u>The PS2 state</u>

$n_+ = 1, \quad n_- = n_2 \leq 1,$

$E(n_+=1) = E_{CDW}(n=1),$

$E(n_-) = E_{NO}(n_-).$

$$E_{PS2}(n) = -\frac{D}{2}\coth\frac{2D}{2\mathbb{W}_Q} + \frac{1}{2}\mathbb{W}_o + (1-n)\left(D(1-n_2) - \mathbb{W}_o n_2\right), \tag{A5}$$

where

$$n_2 = 1 - \left(\frac{\coth\frac{2D}{2\mathbb{W}_Q} - 1}{1 + \frac{\mathbb{W}_o}{D}}\right)^{1/2}.$$

<u>The PS3 state</u>

$(n_+ = 2, n_- = 0)$

$$E_{PS3}(n) = \frac{1}{2}\left(|U| - 2\mathbb{W}_o\right)n, \tag{A6}$$

where

$E_{PS3} < E_{SS}$ for $\quad \frac{\mathbb{W}_o}{D} < \frac{|U|}{2D} - \coth\frac{2D}{|U|}.$

# FIGURE CAPTIONS

**Fig. 1.**

Ground state phase diagram(s) of model (1) with $W_{ij}$ nn, calculated for the rectangular density of states, for several fixed values of $|U|/D$ (numbers to the curves). Dashed lines denote the range of existence of the M phase solution. Left of these lines $E_o^M < E_o^{SS}$ (but $E_o^M > E_o^{PS}$). On the vertical axis, i.e. for n = 1, the ground state is the CDW phase at any $W_{ij} > 0$, whereas for $W_{ij} = 0$ the SS, M and CDW phases are degenerated.

For $W_c < W < 0$ the SS phase is stable, whereas for $W < W_c$ the ground state is the state of electron droplets (PS3).

**Fig. 2.**

Ground state phase diagram of model (1) with the intersite coupling $W_{ij}$ including the nn part $W_1$ ($W_o = zW_1$) and the nnn (second neighbour) part $W_2$, calculated for a) $x = z_2W_2/zW_1 = 0.5$, b) $x = z_2W_2/zW_1 = -0.5$. Dotted line shows the boundary between the PS1 and SS for $x = 0$. Dashed lines denote the range of existence of the M phase solution. $|U|/D = 0.8$.

**Fig. 3.**

Ground state phase diagram of model (1) with the intersite coupling $W_{ij}$ including the nn part $W_1$ ($W_o = zW_1$) and the nnn (second neighbour) part $W_2$, calculated for a) $|U|/D = 0.4$, b) $|U|/D = 0.8$ as a function of $x = z_2W_2/zW_1$. Dashed lines denote the range of existence of the M phase solution.

**Fig. 4.**

Ground state phase diagram of the model (1) with the intersite coupling $W_{ij}$ including the nn part ($W_o = zW_1$) and the long-range Coulomb repulsion $W_{LR} = \sum_{n \geq 2}^{\infty} z_n W_n$, plotted as a function of $W_o/D$ and $|1-n|$ for $x = W_{LR}/W_o = 0.5$

Dashed line shows the boundary between the PS1 and SS for $x = 0$. The M $\leftrightarrow$ SS transition is of $2^{nd}$ order, whereas PS1 $\leftrightarrow$ M and PS1 $\leftrightarrow$ SS are of $1^{st}$ order.
   a) $|U|/D = 0.4$, b) $|U|/D = 0.8$.

**Fig. 5.**

Ground state phase diagram of model (2) (2 $W_{ij} = K_{ij} - J_{ij}$)

a) the case of nn interactions $W_0 = zW_1 \neq 0$, $W_{n\geq 2} = 0$

The PS1 state is strictly degenerated with the M phase ($E_0(PS1) = E_0(M)$) in the whole range of stability of both these states.

b) the case of repulsive nnn interaction:

$z_2W_2/zW_1 = 0.25$ (dashed line indicates the range of existence of the PS1 state solutions)

b) the case of atracttive nnn interaction:

$z_2W_2/zW_1 = -0.25$ (dashed line indicates the range of existence of the M state solutions)

**Table 1.**

Types of ordered states in the systems considered.

| TYPE OF ORDERED STATE | ORDER PARAMETERS |
|---|---|
| 1) SUPERCONDUCTING PHASE **(SS)** | $x_o = \frac{1}{N}\sum_i \langle c_{i\uparrow}^+ c_{i\downarrow}^+ \rangle \neq 0$, model (1),<br>$x_o = \frac{1}{N}\sum_i \langle b_i^+ \rangle \neq 0$, model (2),<br>$(b_i^+ = \rho_i^+)$ |
| 2) CHARGE–ORDERED PHASE **(CDW)** | $n_Q = \frac{1}{N}\sum_{i,\sigma} \langle n_{i\sigma} \rangle e^{i\vec{Q}\vec{R}_i} \neq 0$, model (1),<br>$n_Q = \frac{1}{N}\sum_i \langle b_i^+ b_i \rangle e^{i\vec{Q}\vec{R}_i} \neq 0$, model (2), |
| 3) HOMOGENEOUS MIXED PHASE **(M)** | $n_Q \neq 0$, $x_o \neq 0$,<br>$x_Q = \frac{1}{N}\sum_i \langle c_{i\uparrow} c_{i\downarrow} \rangle\, e^{i\vec{Q}\vec{R}_i} \neq 0$, model (1),<br>$x_Q = \frac{1}{N}\sum_i \langle b_i^+ \rangle\, e^{i\vec{Q}\vec{R}_i} \neq 0$, model (2), |
| 4) NONORDERED PHASE **(NO)** | — |
| 5) PHASE SEPARATED STATE of CDW and SS **(PS1)** | domains of $x_o \neq 0$ ($n_-$)<br>and $n_Q \neq 0$ ($n_+$)<br>($n_+ = 1$, $n_- < 1$) |
| 6) PHASE SEPARATED STATE of CDW and NO **(PS2)** | domains of<br>and $n_Q \neq 0$ ($n_+$)<br>($n_+ = 1$, $n_- < 1$) |
| 7) PHASE SEPARATED STATE of two NO phases – state of electron droplets **(PS3)** | $n_+ - n_- \neq 0$<br>$x_o = x_Q = n_Q = 0$<br>($n_+ > 1$, $n_- < 1$) |

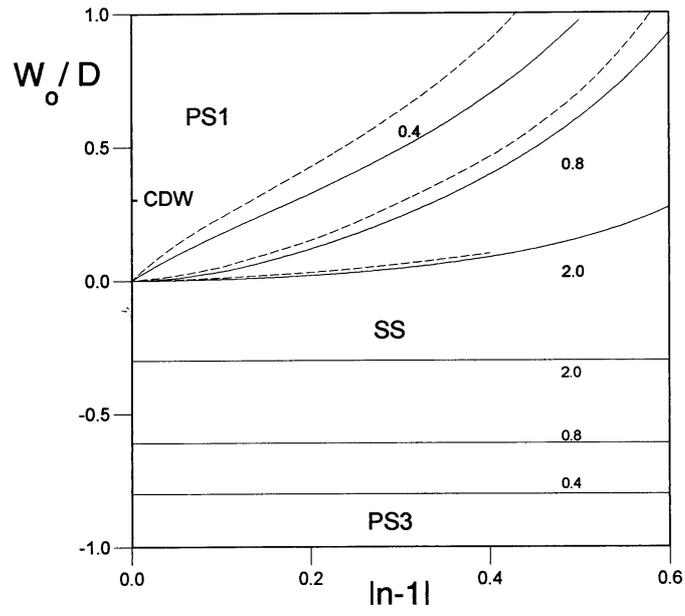

Fig. 1

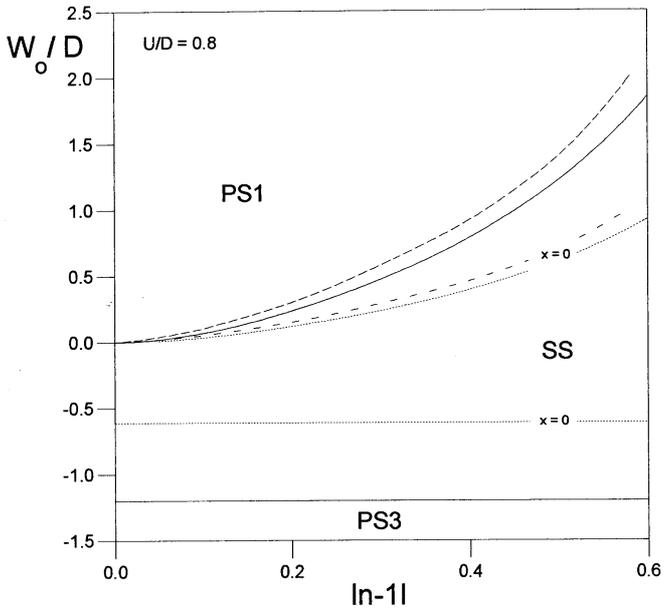

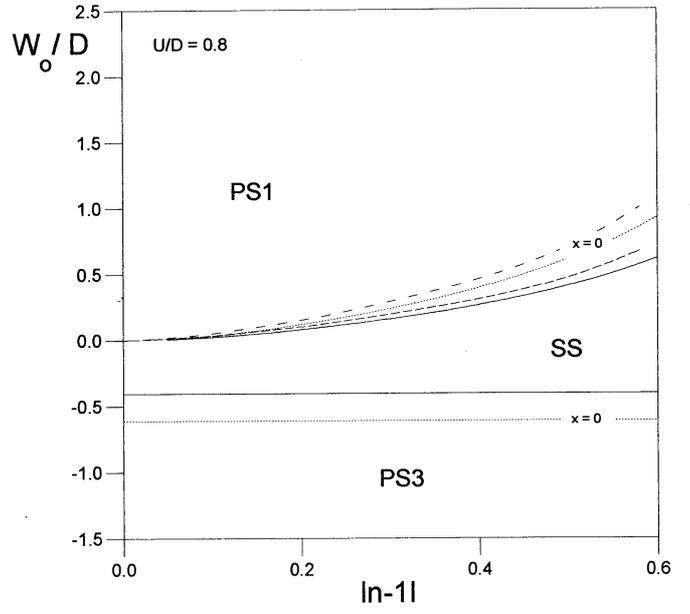

Fig. 2a
Fig. 2b

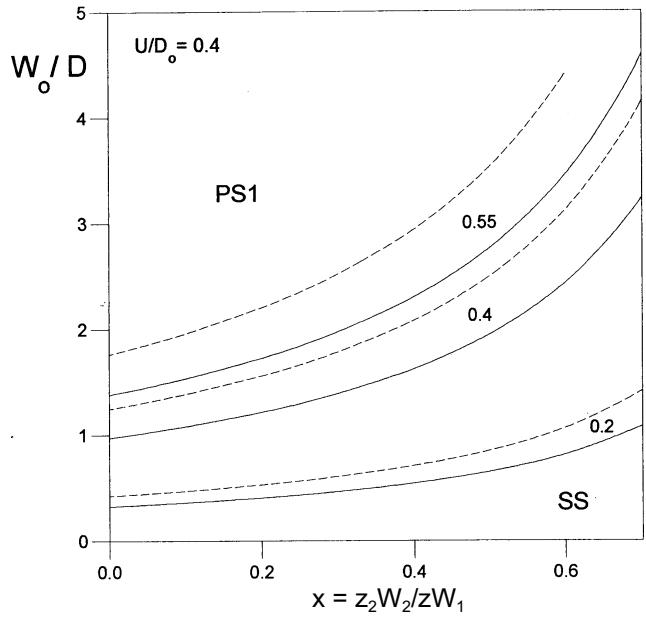

Fig. 3a

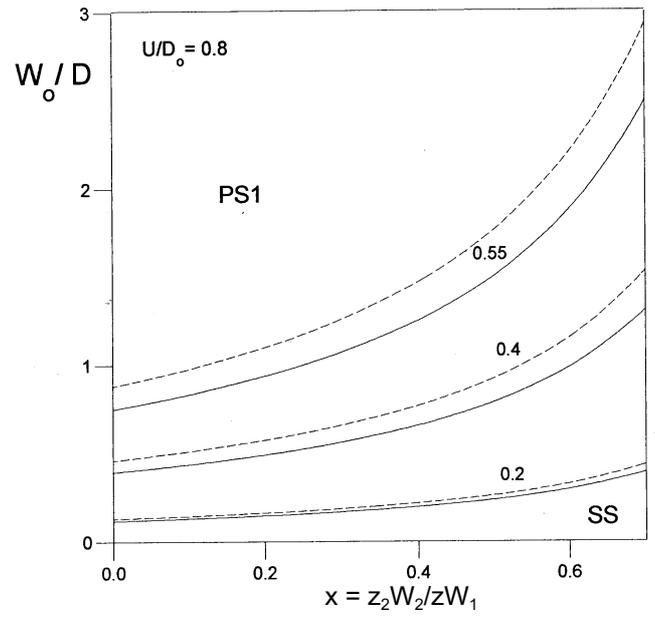

Fig. 3b

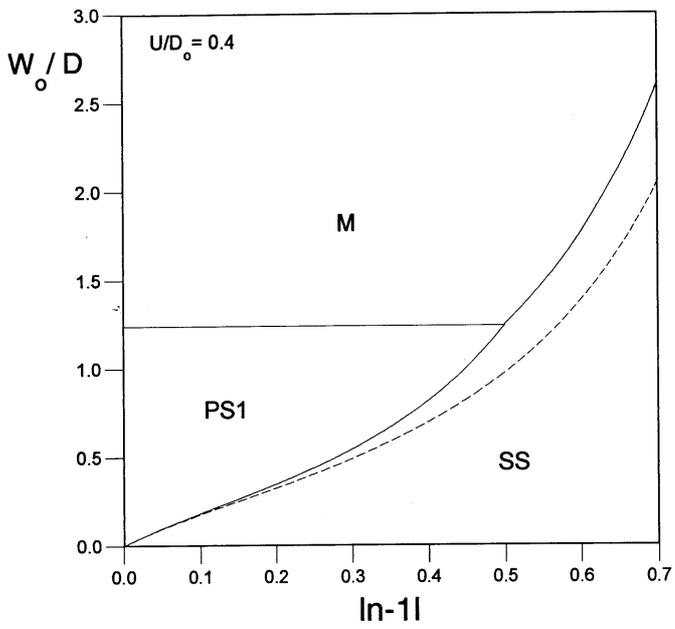

Fig. 4a

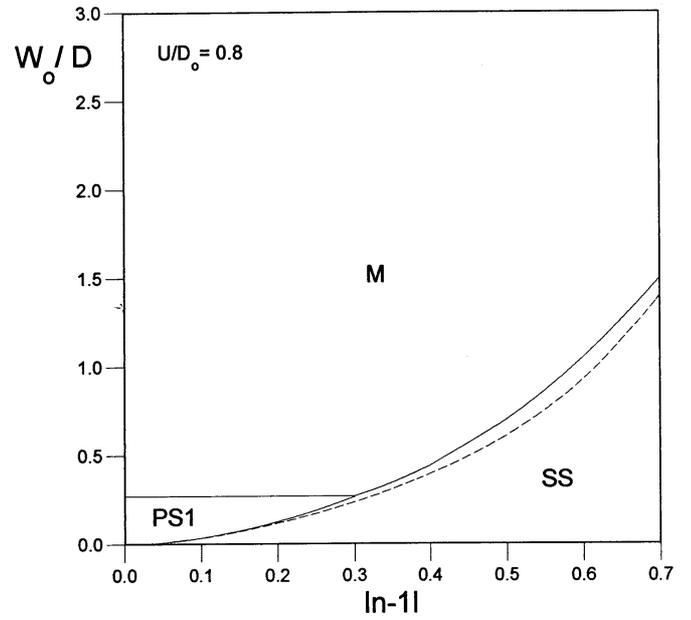

Fig. 4b

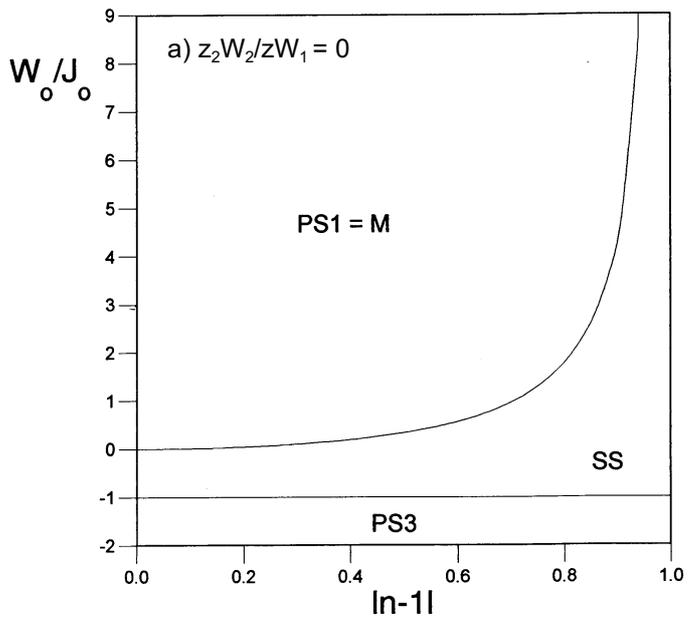

Fig. 5a

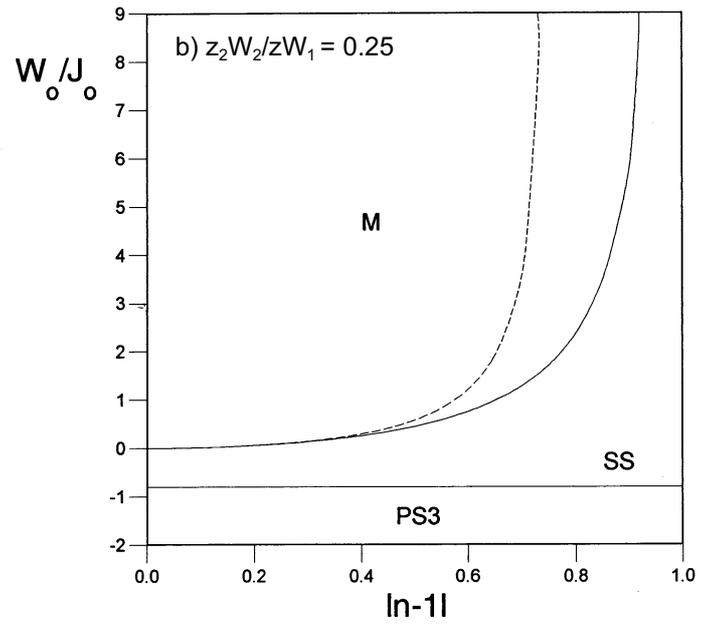

Fig. 5b

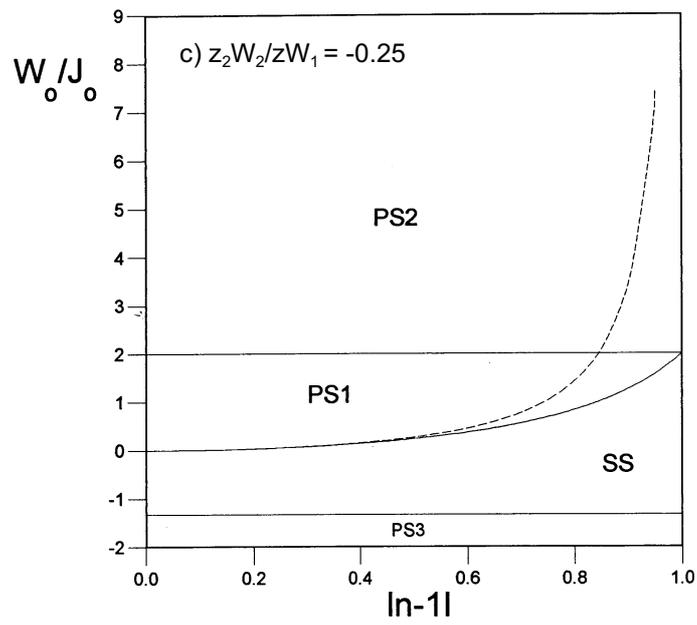

Fig. 5c